\documentclass[fleqn,twoside]{article}
\usepackage{espcrc2}
\usepackage{graphicx}
\newcommand{\AmS}{{\protect\the\textfont2
  A\kern-.1667em\lower.5ex\hbox{M}\kern-.125emS}}

\title{Remarks on the Causality, Unitarity and Supersymmetric Extension of
the Lorentz and CPT-Violating Maxwell-Chern-Simons Model}
\author{A.P. Ba\^{e}ta Scarpelli\address[CBPF]{{\it Centro Brasileiro de 
        Pesquisas F\'{i}sicas (CBPF)}, \\
        Coordena\c{c}\~{a}o de Teoria de Campos e Part\'{i}culas (CCP), \\
        Rua Dr. Xavier Sigaud, 150 - Rio de Janeiro - RJ 22290-180 - Brasil}
		\address[GFT]{{\it Grupo de F\'{i}sica Te\'{o}rica Jos\'{e} Leite Lopes } -
        Petr\'{o}polis - RJ - Brasil}, 
		H. Belich\addressmark[CBPF] \addressmark[GFT], 
		J.L. Boldo\addressmark[GFT] \address{{\it Universidade Federal 
		do Esp\'{i}rito Santo (UFES)},\\
        Departamento de F\'{i}sica e Qu\'{i}mica, Av. Fernando Ferrari, S/N\\
        Goiabeiras, Vit\'{o}ria - ES, 29060-900 - Brasil},
        L.P. Colatto\addressmark[CBPF] \addressmark[GFT],
		J.A. Helay\"{e}l-Neto\addressmark[CBPF] \addressmark[GFT] and 
		A.L.M.A. Nogueira\addressmark[CBPF] \addressmark[GFT]%
        \thanks{{\tt e-mails:scarp\_ufmg@yahoo.com, belich@cbpf.br,
		jboldo@cce.ufes.br, colatto@cbpf.br, helayel@cbpf.br, nogue@cbpf.br}}}        
\begin{document}
\begin{abstract}
\noindent
The gauge-invariant Chern-Simons-type Lorentz- and CPT-breaking term is here
re-assessed and issues like causality, unitarity, spontaneous gauge-symmetry
breaking are investigated. Moreover, we obtain a minimal extension of such a
system to a supersymmetric environment. We comment on resulting peculiar
self-couplings for the gauge sector, as well as on background contribution
for gaugino masses.
\end{abstract}
\maketitle
\section{\ Introduction\-}
\noindent
Lorentz and CPT invariances are cornerstones in modern Quantum Field Theory,
both symmetries being respected by the Standard Model for Particle Physics.
Nevertheless, nowadays one faces the possibility that this scenario is only
an effective theoretical description of a low-energy regime, an assumption
that leads to the idea that these fundamental symmetries could be violated
when one deals with energies close to the Planck scale \cite{Jackiw}. Taking
this viewpoint, several approaches to analyze the violation of Lorentz
symmetry have been proposed in the literature. Eventually a common feature
arises: the violation is implemented by keeping either a four-vector (in a
CPT-odd term \cite{Jackiw}) or a traceless symmetric tensor (CPT-preserving
term \cite{Kostelec1}) unchanged by particle inertial frame transformations 
\cite{Colladay} which is generally called spontaneous violation.
Furthermore, the issue of preserving supersymmetry (Susy) while
violating Lorentz symmetry is addressed to \cite{berger}. This breaking of
Lorentz symmetry is also phenomenologically motivated as a candidate to
explain the patterns observed in the detection of ultra-high energy cosmic
rays, concerning the events with energy above the GZK ( $E_{GZK}\simeq
4\times 10^{19}~eV.T$) cutoff \cite{raios}. Moreover, measurements of radio
emission from distant galaxies and quasars verify that the polarization
vectors of these radiations are not randomly oriented as naturally expected.
This peculiar phenomenon suggests that the space-time intervening between
the source and observer may be exhibiting some sort of optical activity, the
origin of which is not known.

In this context, in Section 2, we analyze the possibility of having
consistency of the quantization of an Abelian theory which incorporates the
Lorentz- and CPT-violating term:
\begin{equation}
\Sigma _{CS}=-\frac{1}{4}\int dx^{4}\epsilon ^{\mu \nu \alpha \beta }c_{\mu
}A_{\nu }F_{\alpha \beta },  \label{jac}
\end{equation}
whenever gauge spontaneous symmetry breaking (SSB) takes place. The analysis
is carried out by pursuing the investigation of unitarity and causality as
read off from the gauge-field propagators. We therefore propose a discussion
at tree-approximation, without going through the canonical quantization
procedure for field operators. In this investigation, we concentrate on the
analysis of the residue matrices at each pole of the propagators. Basically,
we check the positivity of the eigenvalues of the residue matrix associated
to a given simple pole in order that unitarity may be undertaken.
Higher-order poles unavoidably plague the theory with ghosts; this is why
our analysis of the residues is restricted to the case of the simple poles.
We shall find that only for $c_{\mu }$ space-like both causality and
unitarity can be ascertained. On the other hand, considering that SSB is
interesting in such a situation (since the mass generation mechanism induced
by the Higgs scalar presupposes that the theory is Lorentz invariant), it
was showed in the work of Ref. \cite{Baeta} that, once Lorentz symmetry is
violated, there is the possibility of evading this mechanism, such that a
gauge boson mass is not generated even if SSB of the local U(1)-symmetry
takes place.

In view of the interesting features of the original (bosonic) model, many
interesting aspects may show up whenever supersymmety is brought about.
Especially, the fermionic (gaugino) mass generation opens up a relevant
discussion in connection with the presence of background-fermion
condensation. The first proposal of a supersymmetry-preserving Lorentz
violation was carried out in the work of Ref. \cite{berger}. The aim of that
work was to investigate whether one could maintain desired properties of
supersymmetric systems, namely, cancellation of divergences and the patterns
of spontaneous breaking schemes, while violating Lorentz symmetry. A Lorentz
breaking tensor with constants entries has been adopted following an
original suggestion given by Colladay \cite{Colladay}. Working upon a
modified Wess-Zumino model, the authors of Ref. \cite{berger} had
demonstrated that convenient changes of the Susy-algebra of fermionic
charges and of Susy-covariant derivatives expressions were enough to define
a Susy-like invariance for the Lorentz violating starting theory. As a
matter of fact the modification of the algebra was achieved by adding a
particular tensor-dependent central term, of the $k_{\mu \upsilon }\partial
^{\nu }$ type, where $k_{\mu \upsilon }$ exhibits real symmetric traceless
tensor properties. As a net result, it was shown that a model for a
modified-Susy invariant but Lorentz non-invariant {\it matter} system can be
built. Moved by a different perspective, we present, in Section 3, an
analysis on Lorentz and Susy breakings concerning degrees of freedom in the 
{\it gauge} field sector. We carried out the supersymmetric minimal
extension for the Chern-Simons-like term (\ref{jac}), preserving the usual $%
(1+3)$-dimensional Susy algebra \cite{bel}. The breaking of Susy will follow the very
same route to Lorentz breaking: the statement that $c_{\mu }$ is a constant
(in the active sense) vector triggers both Lorentz and, as we shall comment
on, Susy breakings. Choosing appropriate superfield extensions for the background
prevents the model from displaying higher-spin excitations, and
interesting self-couplings for the gauge sector as well as background
contribution for the gaugino masses come up naturally as a consequence of
the (initially) supersymmetric structure. The background-fermion condensates 
play an interesting role for the gaugino mass generation. This shall become
clear after the component-field action is written down. Our conclusions are 
presented in Section 4.

\section{The Gauge-Higgs Model}

We propose to carry out our analysis by starting off from the action: 
\begin{eqnarray}
\Sigma &=&\int d^{4}x\left\{ -\frac{1}{4}F_{\mu \nu }F^{\mu \nu }+ \right.
\nonumber \\%
&&-\left. \frac{\mu }{4%
}v_{\mu }A_{\nu }F_{\kappa \lambda }\varepsilon ^{\mu \nu \kappa \lambda }+ 
\frac{M^{2}}{2}A_{\mu }A^{\mu }\right\} ,
\end{eqnarray}
where $\mu v_{\mu }=c_{\mu }$ and the mass term, $M^{2}$, comes from the
spontaneous symmetry breaking mechanism. The propagator may be obtained
after a lengthy algebraic manipulation. Its explicit form in momentum space
is 
\begin{eqnarray*}
\langle A_{\mu }A_{\nu }\rangle =\frac{i}{D}\left\{ -(k^{2}-M^{2})\left(
\eta _{\mu \nu }-\frac{k_{\mu }k_{\nu }}{k^{2}}\right) \right. + \nonumber \\
+ \left( \frac{D}{M^{2}
}-\frac{\mu ^{2}(v\cdot k)^{2}}{(k^{2}-M^{2})}\right) \frac{k_{\mu }k_{\nu }
}{k^{2}}  + \nonumber \\
 -i \mu \varepsilon _{\mu \nu \kappa \lambda }v^{\kappa }k^{\lambda }-
\frac{\mu ^{2}k^{2}}{(k^{2}-M^{2})}v_{\mu }v_{\nu }+ \nonumber \\ 
+ \left. \frac{\mu ^{2}(v\cdot k)
}{(k^{2}-M^{2})}\left( v_{\mu }k_{\nu }+ 
+ v_{\nu }k_{\mu }\right) \right\} ,
\label{13}
\end{eqnarray*}
where $D(k)=(k^{2}-M^{2})^{2}+\mu ^{2}v^{2}k^{2}-\mu ^{2}(v\cdot k)^{2}.$

The analysis carried out with the help of this propagator reveals that
unitarity is always violated for $v^{\mu }$ time-like and null \cite{Baeta}.
Whenever the external vector is space-like, physically consistent
excitations may be found that present a single degree of freedom each; in
order to prove this, let us check the character of the poles present for $%
v_{\mu }$ space-like. Knowing that 3 different poles show up, we have to go
through the study of the residue matrix of the vector propagator at each of
its poles $k_{0}^{2}=\left( M^{2}+\vec{k}^{2}\right) $, $k_{0}^{2}=m_{\pm
}^{2}$, where $m_{+}$ and $m_{-}$ correspond to the zeroes of $D(k)$.

To infer about the physical nature of the simple poles, we have to calculate
the eigenvalues of the residue matrix for each of these poles. This is done
in the sequel. Before quoting our results, we should say that, without loss
of generality, we fix our external space-like\ vector as given by $v^{\mu
}=(0;0,0,1)$. The momentum propagator, $k^{\mu }$, is actually a
Fourier-integration variable, so we are allowed to pick a representative
momentum whenever $k^{2}>0$. We pursue our analysis of the residues by
taking $k^{\mu }=(k^{0};0,0,k^{3})$.

With $k_{0}^{2}=m_{\pm }^{2}$, we have that 
\begin{eqnarray}
m_{\pm }^{2}&=&\frac{1}{2}\left\{ 2\left( M^{2}+{k_{3}}^{2}\right) + \mu ^{2} \right.
 \nonumber \\
&&\left. \pm \mu \sqrt{\mu
^{2}+4\left( M^{2}+{k_{3}}^{2}\right) } \right\};
\end{eqnarray}
and we find only a single non-vanish eigenvalue (to each pole) for the
residue matrix: 
\begin{equation}
\lambda _{\pm }=\frac{2\left| m_{\pm }\right| }{\sqrt{\mu ^{2}+4\left(
M^{2}+{k_{3}}^{2}\right) }}>0 .
\end{equation}
The calculations above confirm the results found by the authors of Ref. \cite
{Adam}: for a space-like $v^{\mu }$, the pole of $D(k)$ respect causality
(they are not tachyonic) and correspond to physically acceptable $1-$%
particle states with $1$ degree of freedom, since the residue matrix
exhibits a single positive eigenvalue.

Finally, we are left with the consideration of the pole $k_{0}^{2}=\left(
M^{2}+{k_{3}}^{2}\right) $. Here, again, we have obtained only a non-vanish
eigenvalue for the residue matrix: $\lambda =\frac{1}{M^{2}}\left(
M^{2}+2{k_{3}}^{2}\right) >0$. This opens up a very interesting conclusion:
the $M^{2}$-pole, appearing in the longitudinal sector, describes a
physically realizable scalar mode. We are before a very peculiar result: The
vector potential accommodates $3$ physical excitations, each of them
carrying a single degree of freedom; so, the external background influences
the gauge field by drastically changing its physical content: instead of
describing a $3-$degree of freedom massive excitation, it rather describes \ 
$3$ different massive excitations, each carrying one physical degree of
freedom.

\section{The Supersymmetric Extension of the Maxwell-Chern-Simons Model}

Adopting covariant superspace-superfield formulation, we propose the
following minimal extension for (\ref{jac}): 
\begin{equation}
A=\int d^{4}xd^{2}\theta d^{2}\bar{\theta}\left\{ W^{a}(D_{a}V)S+ 
c.c. \right\}  \label{superjac}
\end{equation}
where the superfields $W_{a}$, $V$, $S$ and the Susy-covariant derivatives $%
\ D_{a}$, $\overline{D}_{\dot{a}}$ hold the definitions: 
\begin{eqnarray*}
D_{a} =+\frac{\partial }{\partial \theta ^{a}}+i{\sigma ^{\mu }}_{a\dot{a}}%
\bar{\theta}^{\dot{a}}\partial _{\mu } ,\\
\overline{D}_{\dot{a}} =-\frac{\partial }{\partial \bar{\theta}^{\dot{a}}}%
-i\theta ^{a}{\sigma ^{\mu }}_{a\dot{a}}\partial _{\mu };
\end{eqnarray*}
from ${\overline{D}}_{\dot{b}}W_{a}\left( x,\theta ,\bar{\theta}\right) =0$
and $D^{a}W_{a}\left( x,\theta ,\bar{\theta}\right) =$ $\overline{D}_{\dot{a}%
}\overline{W}^{\dot{a}}\left( x,\theta ,\bar{\theta}\right) $, it follows
that 
\[
W_{a}(x,\theta ,\bar{\theta})=-\frac{1}{4}\overline{D}^{2}D_{a}V.
\]
Its $\theta $-expansion reads as below: 
\begin{eqnarray*}
W_{a}(x,\theta ,\bar{\theta}) =\lambda _{a}\left( x\right) +i\theta ^{b}{%
\sigma ^{\mu }}_{b\dot{a}}\bar{\theta}^{\dot{a}}\partial _{\mu }\lambda
_{a}\left( x\right) + \\
-\frac{1}{4}{\bar{\theta}}^{2}\theta ^{2}\partial^{\mu} \partial _{\mu}\lambda
_{a}\left( x\right)  
+2\theta _{a}D\left( x\right) + \\
-i\theta ^{2}\bar{\theta}^{\dot{a}}{\sigma
^{\mu }}_{a\dot{a}}\partial _{\mu }D\left( x\right) 
+{{{\sigma }^{\mu \nu }}_{a}}^{b}\theta _{b}F_{\mu \nu }\left( x\right) + \\
-
\frac{i}{2}{{{\sigma }^{\mu \nu }}_{a}}^{b}{\sigma ^{\alpha }}_{b\dot{a}%
}\theta ^{2}\overline{\theta }^{\dot{a}}\partial _{\alpha }F_{\mu \nu
}\left( x\right) + 
i{\sigma^{\mu}}_{a\dot{a}}\partial _{\mu }\overline{\lambda }^{%
\dot{a}}\left( x\right) \theta ^{2}
\end{eqnarray*}
and $V=V^{\dagger }$. The Wess-Zumino gauge choice is taken  for $V$
with no loss of generality as far as the action (\ref{superjac}) is
gauge-invariant. The background superfield is so chosen to be a chiral one.
Such a constraint restricts the maximum spin component of the background to
be an $s=\frac{1}{2}$ component-field, showing up as a Susy-partner
for a spinless dimensionless scalar field. Also, one should notice that $S$
happens to be dimensionless. Taking $\overline{D}_{\dot{a}}S\left( x\right)=0$ 
the superfield expansion for $S$ then reads: 
\begin{eqnarray*}
S(x) 
=s(x) +i\theta \sigma ^{\mu }\overline{\theta }\partial
_{\mu }s( x) -\frac{1}{4}{\bar{\theta}}^{2}\theta ^{2} \partial^{\mu} \partial _{\mu}
s( x) + \\
+\sqrt{2}\theta \psi ( x) +\frac{i}{\sqrt{2}}\theta ^{2}%
\overline{\theta }\overline{\sigma }^{\mu }\partial _{\mu }\psi (
x) +\theta ^{2}F( x) .
\end{eqnarray*}
The component-wise counterpart for the action (\ref{superjac}) is as
follows: \bigskip 
\begin{eqnarray}
A_{comp.} =\int d^{4}x \left\{ -\frac{1}{2}(s+s^{\ast })F_{\mu \nu
}F^{\mu \nu }+ \right. \nonumber \\
+ \frac{i}{2}\partial _{\mu }(s-s^{\ast })\varepsilon ^{\mu
\alpha \beta \nu }F_{\alpha \beta }A_{\nu }+4D^{2}(s+s^{\ast })+  
\nonumber \\
-2is\,\lambda \,\sigma ^{\mu }\partial _{\mu }\overline{\lambda }%
-2is^{\ast }\,\overline{\lambda }\,\overline{\sigma }^{\mu }\partial _{\mu
}\lambda + \nonumber \\ 
-\sqrt{2}\lambda (\sigma ^{\mu \nu })F_{\mu \nu }\psi +
\sqrt{2}%
\overline{\lambda }(\overline{\sigma }^{\mu \nu })F_{\mu \nu }\overline{\psi 
}
+ \lambda \,\lambda F+ \nonumber \\ 
\left. +\overline{\lambda }\,\overline{\lambda }%
F^{\ast }-  2\sqrt{2}\lambda \,\psi D+ 
 -2\sqrt{2}\, \overline{\lambda }\,\overline{%
\psi }D \right\}   \label{comps}
\end{eqnarray}
As one can easily recognize, the first two lines display the 4D
Chern-Simons-like term (\ref{jac}), where the vector $c_{\mu }$ is expressed
as the gradient of a real background scalar: $c_{\mu }$ $=$ $\partial _{\mu
}\sigma $, for $s$ $=$ $\xi +i\sigma $. Such a reduction of the vector into
a gradient of a scalar field stems directly from the simultaneous
requirements of both gauge\footnote{%
The gauge invariance of action (\ref{superjac}) will become clearly manifest
in the next section, where we rephrase the supersymmetrization of the 4D
Chern-Simons-like term in a formulation restricted to the chiral
(anti-chiral for the h.c. counterpart) sector of superspace.} and
supersymmetry invariances.

Another interesting feature of this model concerns the presence of
self-couplings for the gauge sector: the fermionic background field, $\psi $%
, triggers the coupling of the gauge boson (through the field-strength) to
the gaugino. Moreover, using the field equation for the gauge auxiliary
field $D$ one arrives at a quartic fermionic fields coupling - $\lambda
\lambda \psi \psi $ -, and the background nature of $\psi $ indicates a
background contribution for the gaugino mass\footnote{%
We shall analyze the propagator structure for the gauge component-fields in
a forthcoming communication. We anticipate that a constant $\psi $
component-field configuration is compatible with the supersymmetry algebra.}.

Concerning the breaking of Lorentz symmetry, realized by assuming $c_{\mu }$ 
$=$ $\partial _{\mu }\sigma $ to be constant under the action of particle
inertial frame transformations, one should observe that such an assumption
implies that the scalar component-field $\sigma $ must be linear in the
coordinates, $\sigma =c_{\mu }x^{\mu }$. As a matter of fact, a linear
dependence on $x^{\mu }$ cannot be implemented by means of a Susy-covariant
constraint (i.e., Susy-covariant derivatives acting on $S$), and, in that
sense, the choice of a rigid $\partial _{\mu }\sigma $ breaks Susy in exact
analogy to the Lorentz breaking scheme adopted. To better establish such a
correspondence, one can consider the choice for constant $\partial _{\mu
}\sigma $ to be accompanied by a constant $\psi $ requirement (and a
constant auxiliary field, $F$, as well\footnote{%
In fact, a constant auxiliary field $F$ is singled out as a susy-invariant
parameter, as far as one deals with a constant $\psi $.}). In this context,
a (passive) Susy-transformation keeps the status of all component-fields
unchanged.

As a first step towards generalizations of the presented minimal
Susy-extension, we have also obtained the following non-polynomial
formulation:
\begin{equation}
A_{\mbox{{\small n-p}}}=\frac{1}{4}\int d^{4}x\left\{ d^{2}\theta \left[ W^{a}W_{a}e^
{(hS)}\right]  +  c.c. \right\} ,  \label{expon}
\end{equation}
whose full component-field expression may be found in the work \cite{bel}. One 
should realize, from the expression above, that a quartic fermion-field 
coupling is already present
at order $h^{2}$ , even if the field equation for the auxiliary field $D$ is
not used to eliminate it. 

\section{Concluding Comments}

Working on the {\it gauge}-field sector of a system with a Lorentz breaking
4D-Chern-Simons-like term, we have been able to derive its minimal
supersymmetric extension. One should already realize the presence of new
couplings induced by the background (passive-)superfield components. The
assumption that the Lorentz breaking is implemented by means of a constant
vector, regarded as a background input, finds its Susy-counterpart in a set
of requirements on the space-time dependence of each component-field of the
background superfield, $S$. A scalar field, $s$, linearly dependent on $%
x^{\mu }$, as well as a constant spinor field, $\psi $, arise as a
consequence of gauge invariance, and these results impose that, eventually,
coupling terms are to be regarded as mass terms. A complete analysis of the
propagator structure for the gauge supermultiplet, both in superspace and in
component-fields, is mandatory, including an interesting study of the
gaugino (background-)induced mass. In terms of components, the explicit breaking
of the Lorentz symmetry becomes manifest through the appearance of a gauge 
field-gaugino mixed propagator induced by the action term that involves the gauge
potential, the gaugino and the background fermion. This is a rather peculiar 
point and, in deriving the full set of propagators, it will become clear that the
gauge field and its fermionic partner will share a common dispersion relation,
for which the background-fermion condensate interferes in competition with the
external vector responsible for the Lorentz breaking.  We 
shall very soon report our efforts in this matter elsewhere.

\section{Acknowledgments}

J.A. H.-N. is grateful to Prof. R. Jackiw for helpful discussions and
suggestions. A.P.B.S. and J.L. B. thanks CNPq for financial support.

\end{document}